\documentclass[%
 reprint,
superscriptaddress,
 amsmath,amssymb,
 aps,
]{revtex4-2}

\usepackage{graphicx}
\usepackage{dcolumn}
\usepackage{bm}
\usepackage{booktabs}
\usepackage{array}
\usepackage{appendix}
\usepackage{tikz}
\usepackage{amsmath, bm}
\usepackage{physics}
\usetikzlibrary{positioning, arrows.meta, calc, arrows.meta,positioning,decorations.pathmorphing,shapes.geometric}

\definecolor{unitaryBG}{HTML}{2E8B57}
\definecolor{stateBG}{HTML}{3A6EA5}     
\definecolor{influenceBG}{HTML}{E08A4A} 
\definecolor{costBG}{HTML}{7F5AC6}      
\definecolor{gradDot}{HTML}{333333}     
\definecolor{legLine}{HTML}{2B2B2B}     
\definecolor{dlegLine}{HTML}{205072}

\usepackage[hidelinks]{hyperref}

\newcommand{\affManchester}{Department of Physics \& Astronomy, University of Manchester, Manchester M13 9PL, United Kingdom}
\newcommand{\affEdinburgh}{School of Physics and Astronomy, University of Edinburgh, Edinburgh EH9 3FD, United Kingdom}
\newcommand{\affUCL}{Department of Physics and Astronomy, University College London, Gower Street, London WC1E 6BT, United Kingdom}
\newcommand{\affSheffield}{School of Mathematical and Physical Sciences, University of Sheffield, Sheffield S3 7RH, United Kingdom}

\begin{document}

\preprint{APS/123-QED}

\title{Efficient optimisation of multi-parameter quantum control protocols for strongly-coupled systems}

\author{Si\^on Meredith}
\affiliation{\affManchester}
\affiliation{Centre for Quantum Science and Engineering, University of Manchester, Manchester M13 9PL, United Kingdom}
\author{Oliver Dudgeon}
\affiliation{\affManchester}
\affiliation{Centre for Quantum Science and Engineering, University of Manchester, Manchester M13 9PL, United Kingdom}
\affiliation{\affEdinburgh}
\author{Wojciech Bukalski}
\affiliation{\affUCL}
\author{\mbox{Alistair J.~Brash}}
\affiliation{\affSheffield}
\author{\mbox{Harry J.~D.~Miller}}
\affiliation{\affManchester}
\affiliation{Centre for Quantum Science and Engineering, University of Manchester, Manchester M13 9PL, United Kingdom}
\author{Thomas J.~Elliott}
\affiliation{\affManchester}
\affiliation{Department of Mathematics, University of Manchester, Manchester M13 9PL, United Kingdom}
\affiliation{Centre for Quantum Science and Engineering, University of Manchester, Manchester M13 9PL, United Kingdom}
\author{Jake Iles-Smith}
\affiliation{\affSheffield}

\date{\today}
             
\begin{abstract}
Achieving high-fidelity control in the presence of strong non-Markovian noise is critical for the optimization of emergent solid-state quantum devices.
We present a highly efficient optimization framework that combines automatic differentiation with the non-Markovian uniTEMPO algorithm, enabling direct gradient-based optimization of complex objective functions. We apply this method to semiconductor quantum dots, optimizing multi-pulse excitation schemes—specifically Swing-UP of a Quantum EmmiteR (SUPER) and Floquet-engineered Two-Photon Excitation (FTPE)—for single- and bi-exciton generation. Our approach yields high preparation fidelities within experimentally accessible parameter regimes. By integrating adiabatic rapid passage (ARP), we systematically enhance both SUPER and FTPE, demonstrating that these optimized protocols consistently outperform standard resonant $\pi$-pulses and two-photon excitation. Notably, this performance gap widens at elevated temperatures, establishing the superior thermal robustness of our optimized multi-pulse strategies for real-world quantum hardware.

\end{abstract}

\maketitle

\section{\label{sec:level1}Introduction}


The control of open quantum systems strongly coupled to environmental degrees of freedom is an issue of both practical and foundational importance \cite{Glaser2015}. As quantum technologies transition from proof-of-concept experiments to scalable architectures, the ability to control quantum systems with high precision in the presence of complex, structured environments has become a central bottleneck. In solid-state systems, the electronic degrees of freedom often interact strongly with the vibrational modes of the surrounding lattice; this interplay between coherent control and non-Markovian dissipation induces complex dynamics that fundamentally degrade the fidelity of target states.

Accurately describing and controlling these dynamics necessitates theoretical frameworks that move beyond traditional master equation treatments \cite{breuer2002theory, mccutcheon2011general, PhysRevB.83.165101}. In the strong-coupling regime, non-Markovian memory effects are prominent; such features cannot be captured by standard weak-coupling approximations or Markovian Lindblad dynamics, which assume an instantaneous environmental response. Consequently, the development of robust control protocols requires non-perturbative methods capable of accounting for the memory of the system-environment interaction.

Semiconductor quantum dots (QDs) provide a prototypical platform where these theoretical challenges become acute.  As leading candidates for on-demand single-photon~\cite{Michler2000, PhysRevLett.98.117402, Liu2018, PhysRevLett.126.233601, Uppu2020, Tomm2021, Ding2025} and entangled-photon sources~\cite{PhysRevLett.87.183601, PhysRevLett.96.130501, Stevenson2006, Dousse2010, PhysRevLett.122.113602}, QDs are primarily limited by decoherence arising from a longitudinal acoustic phonon bath. During coherent driving, phonon-induced thermalization disrupts population transfer, precluding deterministic state preparation. While specialized protocols—such as Swing-UP of the quantum EmitteR population (SUPER)~\cite{PRXQuantum.2.040354}, phonon-assisted preparation~\cite{PhysRevLett.114.137401, PhysRevLett.123.017403, PhysRevB.94.045306}, and Floquet-engineered two-photon excitation (FTPE)~\cite{yan2025robustentangledphotongeneration}—have been developed to overcome these effects, they introduce a significant ``optimization gap." These schemes, while offering the advantage of spectral separation from emission energy, necessitate a delicate and intensive balancing act between pulse area, detuning, and pulse width of multiple pulses.

This optimization gap is compounded by a fundamental computational barrier: as control fields become more sophisticated in an effort to combat decoherence, the numerical cost of optimising the control protocol dramatically increases. Standard master equations are often insufficient to capture the full non-Markovian dynamics of strongly coupled systems, necessitating the use of advanced tensor-network methods like Time-Evolving Matrix Product Operators (TEMPO) \cite{strathearn2018efficient, Jorgenson2019}. However, While these tensor-network methods provide a rigorous description of the physics, their high computational cost has traditionally bottlenecked iterative optimization routines that require thousands of evaluations. To address this, recent frameworks have successfully adapted TEMPO for the optimal control of open quantum systems~\cite{Butler2023, OrtegaTaberner2024ProcessTensor, ortegataberner2026quantumcontrolenvironmentopen, ortegataberner2026qubitresetbornmarkovapproximation}. However, despite their generality, the reliance on manually derived gradients can limit flexibility and become inefficient when scaling to complex control protocols. In lieu of full multi-parameter optimisation, SUPER and FTPE schemes typically require simplifying assumptions and fixed control parameters, restricting exploration to a narrow region of parameter space~\cite{PhysRevB.107.195306, Karli2022SUPER, yan2025robustentangledphotongeneration, PhysRevB.109.245304, Torun2026, piccinini2025excitonbiexcitonpreparationcoherent}. Consequently, the optimisation is effectively confined to a local maximum determined by these fixed assumptions, rather than the true global optimum. 

We bridge this gap by introducing a highly efficient framework for multi-pulse optimisation in the presence of non-Markovian system-environment interactions. By integrating automatic differentiation (AD) \cite{baydin2018automaticdifferentiationmachinelearning, innes2019dontunrolladjointdifferentiating} with the non-Markovian uniTEMPO \cite{link2024open,kahlertSimulatingLandauZener2024,sonnerSemigroupInfluenceMatrices2025} algorithm, we enable the calculation of exact gradients for state fidelity across arbitrary parameter dimensions. This allows us to navigate complex control landscapes with unprecedented speed and accuracy, without the need for manually calculating gradients~\cite{Butler2023, OrtegaTaberner2024ProcessTensor, ortegataberner2026quantumcontrolenvironmentopen, ortegataberner2026qubitresetbornmarkovapproximation}. We demonstrate the versatility of this framework by optimising single- and bi-exciton preparation in semiconductor QDs while accounting for numerically exact electron-phonon dynamics. Our approach efficiently handles high-dimensional spaces—optimizing in excess of 23 independent control parameters—to find pulse sequences that remain robust at elevated temperatures. Even under realistic experimental constraints, these protocols significantly outperform standard resonant $\pi$-pulses \cite{Stievater2001, PhysRevLett.87.246401} and two-photon excitation (TPE) \cite{PhysRevB.73.125304} for single- and bi-exciton preparation respectively. 

\section{Differentiable \lowercase{uni}TEMPO Optimisation}
\label{sec:Optimisation}
\begin{figure*}[t!]
\includegraphics[width=\textwidth]{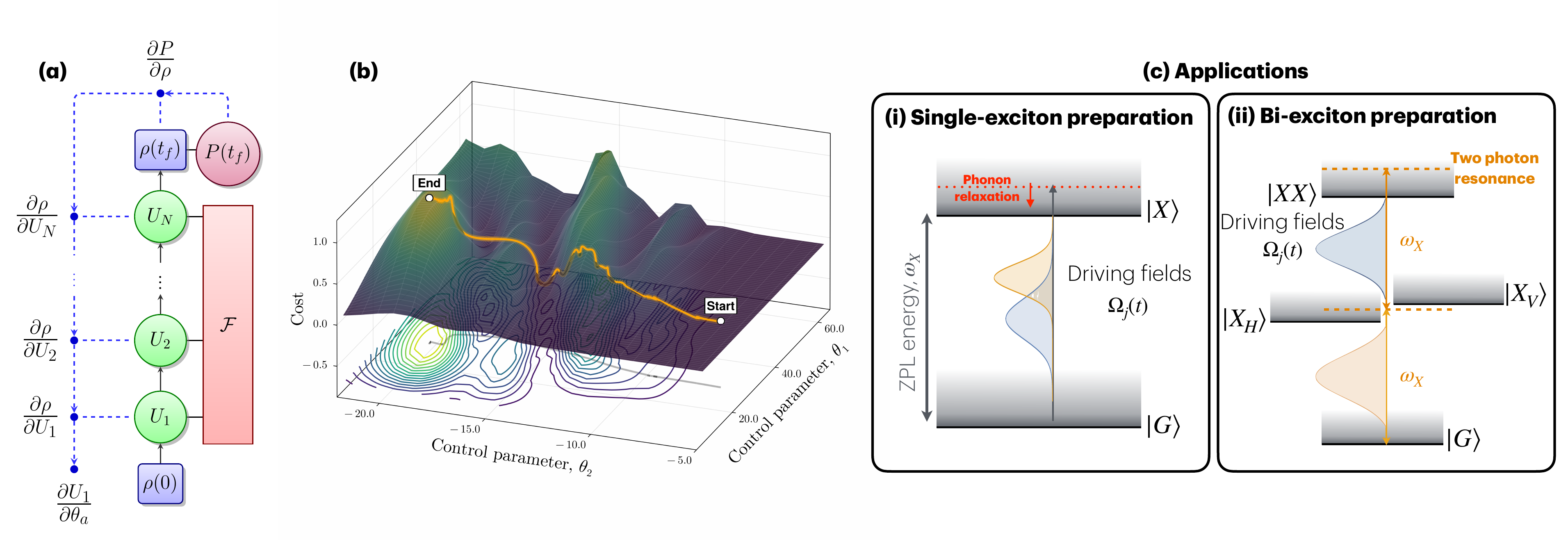}
\caption{(a) Tensor network representation of the cost function $P_X(t_f)$ and the computation of its partial derivative $\partial P_X(t_f)/\partial \theta_a$ with respect to a control parameter $\theta_a$. The network explicitly encodes the multilinear structure of the propagators $\{\mathcal{U}_n\}_{n=1}^N$ and incorporates the influence functional $\mathcal{F}$. Reverse-mode automatic differentiation (backpropagation) is illustrated via adjoint signal propagation through the network. 
(b) 3D projection of the cost landscape $P_X(t_f)$ depicting the optimisation trajectory. The path traces the evolution from the initial parameter configuration (Start) to the converged local maximum (End), guided by L-BFGS. (c) Energy level schematics for (i) single-exciton and (ii) bi-exciton systems. The diagrams highlight the phonon-induced broadening of the transitions and the specific optical driving protocols employed for state preparation.}
\label{fig:AUTO-PT}
\end{figure*}

In general, we aim to maximise a cost function $C$, defined as a functional of the system's reduced state. This cost depends on $N_p$ independent control parameters, which we collectively denote by $\boldsymbol{\theta}$. The resulting optimisation problem defines a high-dimensional landscape, with dimensionality scaling linearly with $N_p$, and quickly becomes intractable for manual or grid-based approaches. Access to the exact gradient $\nabla C$ significantly accelerates numerical convergence, making it highly advantageous for efficiently identifying optimal control sequences.

Evaluating $C$ and its gradients in the strong-coupling regime is highly non-trivial, as standard master-equation approaches fail to capture the necessary memory effects. To accurately describe the system--environment dynamics, we employ uniTEMPO~\cite{link2024open}, a tensor-network method that provides a numerically exact treatment of non-Markovian dissipation. Within this framework, the full environmental influence is encoded in a discrete-time tensor network known as a \textit{process tensor}, which can be computed independently of the control fields acting on the system. The final reduced state is then obtained by iteratively contracting a sequence of parameterised system propagators, $U_n(\boldsymbol{\theta})$, with this tensor network, where $n$ labels the timestep. We refer the reader to Appendix~\ref{appendix:uniTEMPO_algorithm} for further details of the uniTEMPO method. A key advantage of this approach is that the full tensor-network contraction naturally defines a computational graph, enabling the use of automatic differentiation (AD)~\cite{baydin2018automaticdifferentiationmachinelearning, innes2019dontunrolladjointdifferentiating}. 

In this work, we focus on preparing a target state $|\Psi\rangle$ at a final time $t_f$, and therefore maximise the preparation fidelity,
\begin{equation}
P_\Psi(t_f) = \langle \Psi | \rho_S(t_f) | \Psi \rangle.
\label{eq:Cost_func_general}
\end{equation}
This formulation enables efficient evaluation of the gradient $\nabla_{\boldsymbol{\theta}} P_\Psi(t_f)$ with respect to all control parameters,
\begin{equation}
\frac{\partial P_\Psi(t_f)}{\partial \theta_a} 
= \sum_{n=1}^{N} \sum_{i,j,k,l}^{d^2} 
\frac{\partial P_\Psi(t_f)}{\partial \rho_f^i}
\frac{\partial \rho_f^i}{\partial U_n^{jkl}}
\frac{\partial U_n^{jkl}}{\partial \theta_a},
\end{equation}
where $\theta_a \in \boldsymbol{\theta}$ denotes an individual control parameter. Here, the indices correspond to Liouville-space notation, in which density operators are represented as vectors of length $d^2$, with components $\rho^i(t)$, where $d$ is the Hilbert space dimension. Further details on automatic differentiation and its integration into uniTEMPO are provided in Appendix~\ref{appendix:Auto_diff}.
This fully differentiable framework—illustrated diagrammatically in Fig.~\ref{fig:AUTO-PT}(a)—provides direct access to exact gradients, enabling the deployment of advanced numerical solvers to maximize Eq.~\eqref{eq:Cost_func_general}. While the framework is compatible with a variety of gradient-based optimizers, we employ the L-BFGS algorithm~\cite{1995SJSC...16.1190B, bollapragada2018progressivebatchinglbfgsmethod}, which we found consistently outperforms first-order optimisers such as Adam~\cite{kingma2017adammethodstochasticoptimization} and stochastic gradient descent~\cite{ruder2017overviewgradientdescentoptimization} across our target control landscapes.
To mitigate the risk of the optimizer becoming trapped in sub-optimal local maxima, we initialize the L-BFGS routine using a gradient-free warm-up phase. We utilize random search~\cite{zhang2019derivativefreeglobaloptimizationalgorithms} for standard protocols and a differential evolution algorithm~\cite{storn1997differential} for traversing more challenging, highly structured parameter spaces.
Crucially, integrating this optimization routine with the process tensor formalism provides a significant computational advantage over traditional iterative master-equation techniques. Because the discrete-time process tensor depends strictly on the bath parameters and is entirely independent of the coherent control fields applied to the system, the computationally expensive environmental influence need only be calculated once. It is then cached and reused across the thousands of evaluation steps required for the optimization, drastically reducing the overall computational overhead. 
  
While gradient-based optimization has been previously applied to process tensors using manual algebraic derivatives~\cite{Butler2023, OrtegaTaberner2024ProcessTensor,  ortegataberner2026quantumcontrolenvironmentopen, ortegataberner2026qubitresetbornmarkovapproximation}, such approaches require bespoke analytic calculations for every new control Hamiltonian. In contrast, our integration of AD with the uniTEMPO algorithm provides a more flexible, model-agnostic framework. By automating the gradient calculation through the entire simulation, we eliminate the need for manual derivatives, enabling the rapid and efficient optimization of complex quantum protocols across arbitrary parameter spaces, multi-time object functions, and scenarios in which multiple environments are present.

\section{Optimising Quantum State Preparation Schemes}
\label{sec:Theory_prep_schemes}

The optimisation scheme discussed above is a general tool for developing quantum control protocols beyond unitary and Markovian dynamics. 
In this Section, we apply this protocol to the problem of quantum state preparation in solid-state quantum emitters, specifically, the preparation of single- and bi-exciton states in self-assembled QDs.

\subsection{Phonon interactions}
\label{sec:Phonon interactions}

The electronic states of self-assembled QDs couple strongly to longitudinal acoustic (LA) phonons, which dictate the limits of state preparation. This interaction presents a dual challenge: phonon-induced decoherence during driving prevents full population inversion, yet phonon-mediated thermalization—leveraged through off-resonant excitation—offers a robust pathway to the target state. Consequently, integrating a non-perturbative treatment of the electron-phonon interaction into the optimization loop is essential to bridging the gap between theoretical protocols and experimental high-fidelity control.   

We model the quantum dot as an effective three-level system with basis states $\{\ket{G}, \ket{X}, \ket{XX}\}$, representing the ground state, the single-exciton state, and the bi-exciton state, respectively. The total system-environment Hamiltonian is given by $H(t) = H_S(t) + H_E + H_I$, where $H_S(t)$ describes the coherent driving pulses required for state preparation. The longitudinal acoustic phonon bath is described by the free Hamiltonian
\begin{equation}
H_E = \sum_k \omega_k b^\dagger_k b_k,
\label{eq:EnvironmentHamiltonian}
\end{equation}
where $b_k^\dagger$ ($b_k$) are the creation (annihilation) operators for a phonon mode with frequency $\omega_k$ (taking $\hbar=1$).

At low temperatures, the dominant electron-phonon interaction can be described by the linear Hamiltonian,
\begin{equation}
H_I = \sum_{k, \nu} n_\nu\dyad{\nu} (g_k b^\dagger_k + g^*_k b_k),
\label{eq:InteractionHamiltonian}
\end{equation}
where $\lvert \nu \rangle$ denotes the excitonic states and $n_\nu$ is the corresponding number of excitons present. 
The electron-phonon coupling strengths, $g_k$, are well characterised by the super-Ohmic spectral density $J(\omega) = \sum_k |g_k|^2 \delta(\omega - \omega_k) = \alpha \omega^3 e^{-(\omega/\omega_c)^2},$ where we take the coupling strength is $\alpha = 0.027~\text{ps}^2$ and cutoff frequency $\omega_c = 2.2~\text{ps}^{-1}$~\cite{Nazir_2016, Denning:20}. 
This coupling leads to an energetic shift of the system transition energies for each state $\ket{\nu}$, which takes the analytic form,
\begin{equation}
R_\nu =  n_\nu^2\int_{0}^{+\infty} d\omega \frac{J(\omega)}{\omega} = \frac{\sqrt{\pi}}{4} \alpha n_\nu^2\omega_c^3.
\label{eq:PolaronShift}
\end{equation}
Throughout our optimizations, we assume a strict separation of timescales between the ultrafast optical excitation dynamics and the spontaneous emission of the exciton, allowing us to neglect Markovian radiative decay. While this approximation is well-justified for quantum dots in bulk materials or within weakly Purcell-enhanced microcavities, it is not a fundamental restriction of our numerical framework. Incorporating simultaneous Markovian decay channels is entirely feasible, though it would necessitate a modified objective function to properly account for the transient nature of the target state.

\subsection{Single-Excitons}
\label{sec:Single-Excitons}
 For single-exciton generation, we restrict the system to the two-level subspace spanned by the ground state and the single exciton, $\{ \lvert G \rangle, \lvert X \rangle \}$ [see Fig.~\ref{fig:AUTO-PT}(ci)], seeking to optimise the final exciton population $P_X(t_f)$. 
 Working in a frame rotating at the frequency $\omega_X - R_X$, the system Hamiltonian is given by
\begin{equation}
 H_S(t) = R_X \lvert X \rangle \langle X \rvert + \frac{1}{2}\sum_{j=1}^{N} \left[ \Omega_j(t - \tau_j)e^{-i\delta_j t} \sigma^{\dagger} + \text{h.c.} \right],
\label{eq:system Hamiltonian} 
\end{equation}
where $\sigma^\dagger = \lvert X \rangle \langle G \rvert$ ($\sigma = \lvert G \rangle \langle X \rvert$) is the raising (lowering) operator of the quantum dot. The external driving term serves as the controllable element in our framework, comprising a sequence of $N$ independent optical pulses. Each pulse $j$ is defined by its detuning $\delta_j = \omega_j - \omega_X + R_X$, a relative temporal delay $\tau_j$ (with $\tau_1 = 0$), and a Gaussian temporal envelope:
 \begin{equation}
 \Omega_j(t - \tau_j) = \frac{\Theta_j}{\sigma_j\sqrt{2\pi}} e^{-(t - \tau_j)^2/(2\sigma_j^2)},
 \label{eq:Pulse_envelope}
 \end{equation}
where $\Theta_j$ dictates the pulse area and $\sigma_j$ represents the pulse width.
Consequently, any $N$-pulse excitation scheme requires the simultaneous optimization over a high dimensional vector $\mathbf{\Theta} = \{\Theta_j, \sigma_j, \delta_j, \tau_j\}_{j=1}^N$, resulting in a landscape of $4N-1$ independent dimensions.
For standard finite-difference methods, evaluating the gradient over this landscape requires a number of expensive non-Markovian simulations that scales linearly with the parameter count. By contrast, our AD-uniTEMPO framework evaluates the full exact gradient in a single backward pass. This effectively decouples the gradient computation time from the dimensionality of the control space, making the optimization of complex multi-pulse sequences both tractable and highly efficient.

\subsection{Biexciton preparation}
\label{subsec:Bi-Excitons}

For biexciton preparation, the four-level structure of the QD consists of the ground state $|G\rangle$, two linearly polarized exciton branches $|X_H\rangle$ and $|X_V\rangle$, and the biexciton state $|XX\rangle$, as illustrated in Fig.~\ref{fig:AUTO-PT}(cii). However, because the driving lasers are assumed to be vertically polarized throughout, the horizontally polarized exciton state $|X_H\rangle$ remains uncoupled from the excitation dynamics. The system can therefore be reduced to an effective three-level basis $\{|G\rangle, |X\rangle, |XX\rangle\}$, where $|X\rangle$ denotes the vertically polarized exciton state $|X_V\rangle$. We take the objective function to be the final biexciton population $P_{XX}(t_f)$. The biexciton energy is defined as $E_{XX} = 2E_X - E_b$, where $E_X$ is the single-exciton energy and $E_b = 4.28$ ps$^{-1}$ is the biexciton binding energy \cite{yan2025robustentangledphotongeneration}. By moving into a frame rotating with frequency of $(E_{XX}-R_{XX})/2$ we can define the control Hamiltonian for biexciton preperation as 
\begin{equation}
\begin{split}
H_s(t) &= \frac{1}{2}(E_b + R_{XX})\dyad{X} + \frac{R_{XX}}{2} \dyad{XX} \\
&\quad + \frac{1}{2}\sum_{j=1}^{N} \Big[ \Omega_j(t-\tau_j)e^{-i\delta_j t}(\sigma_X^{\dagger}+\sigma_{XX}^{\dagger}) + \mathrm{h.c.}\Big].
\end{split}
\label{eq:H_S_BX}
\end{equation}
Here, the raising operators for the sequential transitions are defined as $\sigma_X^{\dagger} = \lvert X \rangle \langle G \rvert$ and $\sigma_{XX}^{\dagger} = \lvert XX \rangle \langle X \rvert$. The driving pulse envelopes $\Omega_j(t-\tau_j)$ retain the Gaussian profile established in Eq~\eqref{eq:Pulse_envelope}, while the effective detuning of each pulse is now given by $\delta_j = \omega_j - (E_{XX}-R_{XX})/2$. As with the single-exciton case, an $N$-pulse protocol relies on a high-dimensional control space comprising $4N-1$ independent parameters. However, the expansion to a three-level system significantly increases the baseline computational cost of each underlying non-Markovian forward simulation.

\subsection{Chirped Pulses}
\label{sec:Chirped Pulses}

In addition to the baseline parameter set $\{\Theta_j, \sigma_j, \delta_j, \tau_j\}$, we expand the optimization space to include linear frequency chirp. Chirped pulses are the foundation of Adiabatic Rapid Passage (ARP) protocols~\cite{PhysRevB.95.241306, PhysRevLett.106.166801, PhysRevLett.106.067401}, which have been widely utilized for high-fidelity exciton preparation in quantum dots. By adiabatically sweeping the driving frequency, ARP facilitates robust population transfer with reduced phonon-induced dephasing at lower pulse areas compared to standard Rabi oscillations. Motivated by these advantages, we incorporate chirping into our multi-pulse framework. This modification implies that our detunings become linearly time-dependent~\cite{PhysRevB.95.241306}:
\begin{equation}
\delta_j(t) = \delta_{0,j} + \frac{a}{2}t
\label{eq:detuning}
\end{equation}
where $\delta_{0, j}$ is the initial detuning of the $j$-th pulse from resonance and $a$ is the chirp rate describing the rate of frequency change over time. The pulse envelope is modified by the inclusion of a chirp, taking the form 
\begin{equation}
\Omega_j (t) = \frac{\Theta_j}{\sqrt{2 \pi \sigma_{0,j} \sigma_j }} \exp \left( -\frac{(t-\tau_j)^2}{2\sigma_j^2} \right)
\label{eq:Chirp_envelope}
\end{equation}
The parameter $\sigma_{0,j}$ denotes the initial (unchirped) pulse width and $\sigma_j$ is the effective pulse width, essentially the width of the pulse under the influence of the frequency sweep applied due to the chirp. The chirp rate $a$ and the effective pulse width can be described in terms of the chirp coefficient $\lambda_j$, and $\sigma_{0,j}$. The effective pulse width is given by $\sigma_j = \sqrt{\sigma_{0,j}^2 + \lambda_j^2/\sigma_{0,j}^2}$ and $a = d\omega/dt = \lambda_j/(\lambda_j^2 + \sigma_{0,j}^4)$. 

The system Hamiltonian is updated by replacing the Gaussian envelopes and constant detunings with those defined in equations~\eqref{eq:detuning} and~\eqref{eq:Chirp_envelope}. Consequently, this introduces only one additional optimization parameter per pulse, the chirp coefficient $\lambda_j$.

\section{Constrained Optimisation of Multipulse excitation protocols}

A key motivation of this work is not only to demonstrate the power of combining AD and uniTEMPO for multiparameter optimisation, but also to enhance the experimental feasibility of implementing multi-pulse excitation schemes. In particular, our approach enables efficient exploration of the full control landscape within experimentally viable bounds. From an experimental perspective, high pulse energies can drive the system beyond the intended two- or three-level manifold. Furthermore, recent experimental evidence has revealed pronounced bunching on microsecond timescales under SUPER excitation~\cite{piccinini2025excitonbiexcitonpreparationcoherent}, which increases with higher excitation power; this suggests that high-power SUPER pulses induce fluctuations in the charge environment. Strong driving therefore not only introduces unwanted excitations within the semiconductor system, but can also compromise the stability of the emitter itself, further limiting the reproducibility of experimental results.

Despite these challenges, high-fidelity exciton and biexciton preparation has been experimentally demonstrated using multi-pulse excitation schemes. For exciton generation using the SUPER protocol, fidelities reaching an upper bound of $80\%$ have been experimentally observed with pulse areas ranging between $8\pi$ and
$14\pi$~\cite{Boos2024SwingUpQD}. Meanwhile biexciton preparation via FTPE has been reported with an experimental fidelity of $96.1\%$ to the biexciton state for pulse areas exceeding $14\pi$~\cite{yan2025robustentangledphotongeneration}. To address these practical limitations, we incorporate experimental constraints directly into the optimisation framework. Specifically, we restrict the driving fields to a maximum pulse area of $12\pi$ for single-exciton generation and $15\pi$ for biexciton preparation. In addition, we augment the cost function with a spectral penalty that constrains the overlap of the driving pulses with the emitter's zero-phonon line (ZPL) to below 10\%. This reduces resonant scattering of the pump laser and alleviates the need for stringent filtering in photon detection setups. A detailed formulation of the cost function is provided in Appendix~\ref{appendix:The Cost Function}.

\subsection{Optimal two-pulse excitation schemes}

Using the experimental constraints defined above, we consider the case of optimising two pulses for both the single- and bi-exciton preparation using the AD-uniTEMPO framework.
We consider an initial bath temperature of $T=4\mathrm{K}$ and choose a final time $t_f=30$ps. We assume a seperation of timescales between the driving of the emitter and the radiative recombination of the QD. While this may not hold for strongly Purcell enhanced quantum emitters~\cite{Liu2018, Rickert2024}, it would not change the optimisation framework significantly, where instead of the final state population, the cost function would be written in terms of the integrated intensity. For the standard two-pulse SUPER protocol, we obtain a single-exciton state fidelity of $P_X(t_f) = 99.63\%$. The optimization yields pulse areas of $\Theta_1 = 11.84\pi$ and $\Theta_2 = 8.44\pi$, with corresponding detunings of $\delta_1 = -5.36\,\text{ps}^{-1}$ and $\delta_2 = -20.00\,\text{ps}^{-1}$. Both pulses have widths of $\sigma = 1.00\,\text{ps}$ and are separated by a minimal temporal delay of $\tau = 1.00\,\text{ps}$. Similarly, the FTPE-like two-pulse sequence achieves a biexciton preparation fidelity of $P_{XX}(t_f) = 99.29\%$. The optimization yields pulse areas of $\Theta_1 = 14.00\pi$ and $\Theta_2 = 14.04\pi$, with corresponding detunings of $\delta_1 = -13.77\,\text{ps}^{-1}$ and $\delta_2 = 13.93\,\text{ps}^{-1}$. Both pulses again have widths of $\sigma = 1.00\,\text{ps}$ and are separated by a minimal temporal delay of $\tau = -0.15\,\text{ps}$.

\begin{figure}[t!]
\centering
    \hspace{-0.60cm}
    \includegraphics[width=1.05\columnwidth]{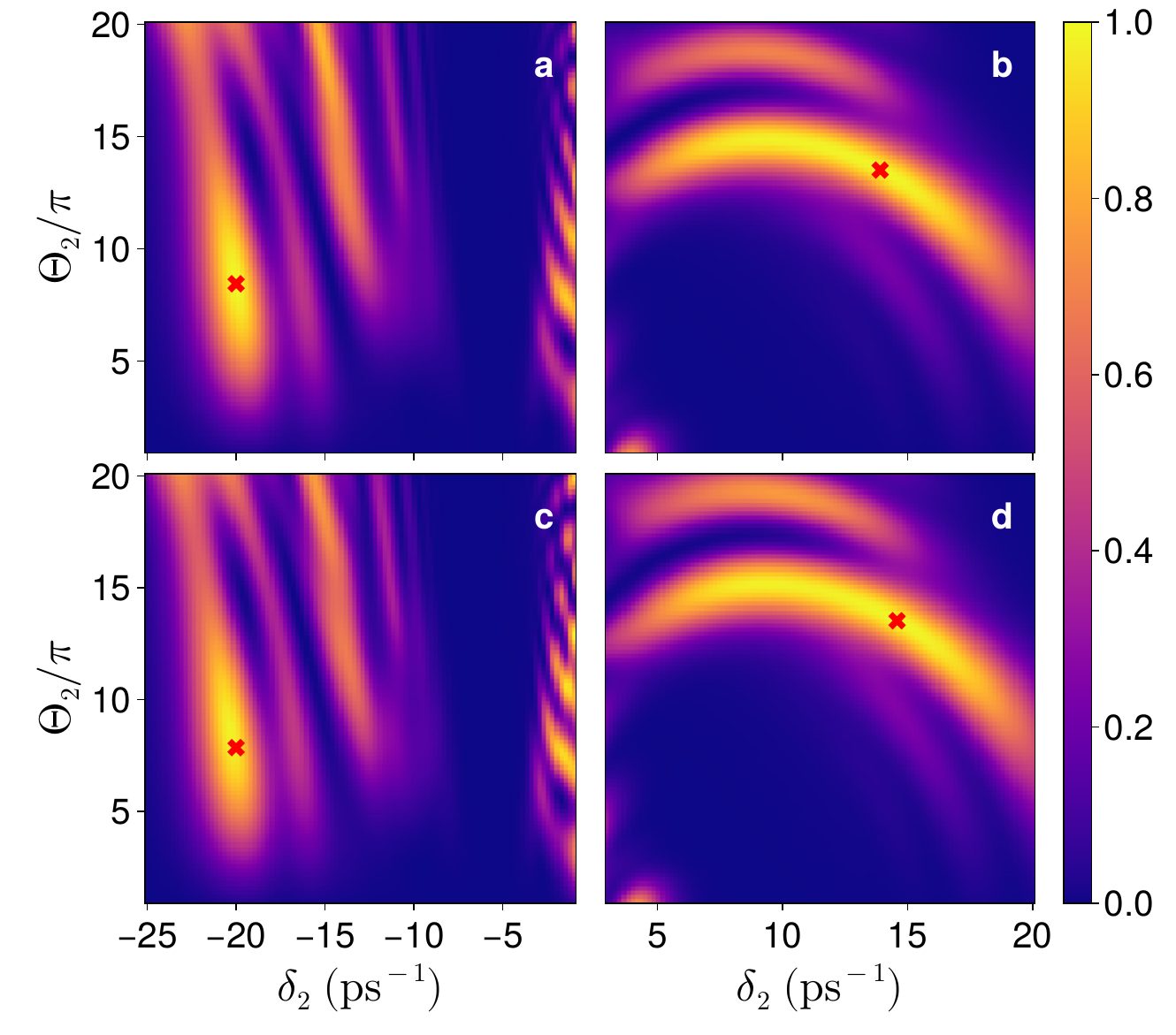}
    \caption{Excited state population as a function of pulse area $\Theta_2$ and detuning $\delta_2$, with other parameters held at their optimized values. Panels (a) and (b) show single- and bi-exciton generation respectively without chirping, while (c) and (d) depict the corresponding cases with chirping included.}
    \label{fig:Contour population}
\end{figure}

To demonstrate the effectiveness of the optimisation, we plot the landscape population inversions as a function of $\Theta_2$ and $\delta_2$ while fixing all other pulse parameters to their optimised values. The result of the full AD-uniTEMPO optimisation are given by the red cross, demonstrating that the optimization scheme has successfully identified the optimal value in the constrained parameter space. It is notable that the maximum achievable single-exciton fidelity marginally exceeds that of the bi-exciton. This discrepancy is rooted in the physical dynamics of the distinct excitation pathways. Single-exciton generation operates within a relatively straightforward two-level subspace. In contrast, bi-exciton preparation relies on a two-photon transition traversing a three-level ladder. Consequently, the intermediate single-exciton state is transiently populated and remains susceptible to phonon-induced dephasing throughout the protocol. These additional decoherence channels inherent to the three-level system seemingly bound the upper limit of the biexciton fidelity compared to the two-level case. To explore whether this environmentally induced decoherence can be further suppressed, we expand the optimization space to include linear pulse chirping. Because our fully differentiable framework computes exact gradients simultaneously across all control parameters, incorporating this additional degree of freedom incurs negligible computational overhead. The addition of chirped pulses yields marginal but consistent gains, refining the state preparation fidelities to $P_X(t_f)$ = 99.66\% and $P_{XX}(t_f)$ = 99.34\%. While the increase in the population inversion is modest, the adiabatic frequency sweep offers a higher mitigation against the effects of increasing temperature, see Section~\ref{sec:temp_dep} for more details.

\subsection{Multi-Pulse scaling and dimensionality}

To systematically probe the limits of our target state preparation, we extend our analysis beyond the standard two-pulse protocol, demonstrating the scalability of our optimization scheme across highly parameterized landscapes. 
This extension is straightforward:
because the analytical gradient computation is fully automated~\cite{baydin2018automaticdifferentiationmachinelearning, innes2019dontunrolladjointdifferentiating} and the non-Markovian environmental influence is cached via the uniTEMPO process tensors~\cite{link2024open}, only the system Hamiltonian requires modification.
This highlights the minimal change required for our optimisation scheme when applied to different control problems, provided that the spectral density and environmental coupling remain unchanged. Crucially, as the number of pulses $N_p$ increases, the dimensionality of the control space expands as $N_\theta = 4N_p - 1$. While optimizing such a high-dimensional space would be computationally prohibitive for standard grid-search or finite-difference methods, our AD-driven approach traverses it efficiently.

\begin{figure}[t!]
\centering
\hspace{-1cm}
    \includegraphics[width=\columnwidth]{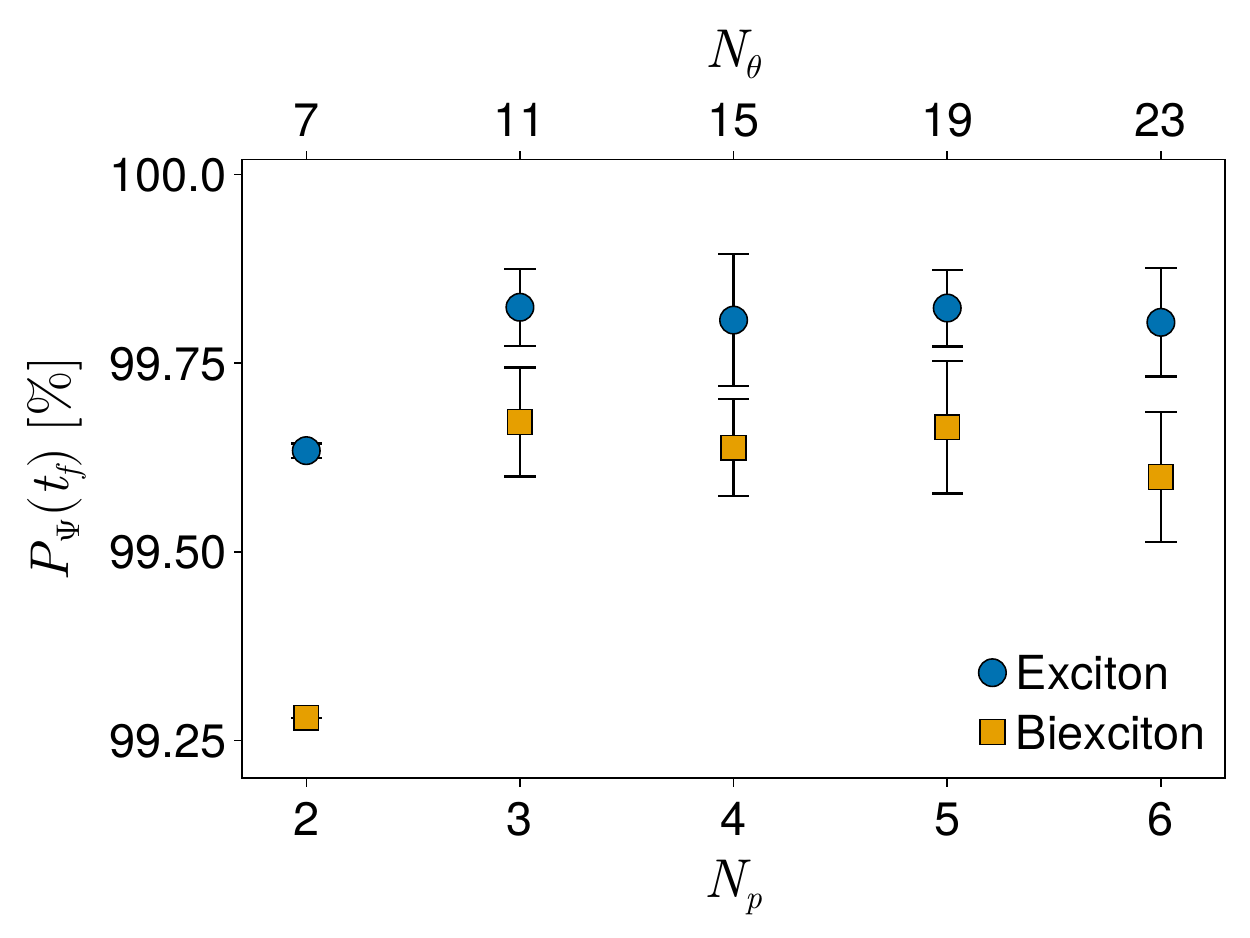}
    \caption{Fidelity of single- and bi-exciton preparation as a function of the number of pulses. $N_{\theta}$ denotes the number of parameters optimized for the protocol, while $N_{p}$ indicates the total number of pulses applied.}
    \label{fig:Multi-pulse-plot}
\end{figure}

Figure~\ref{fig:Multi-pulse-plot} presents the average state-preparation fidelity for both single- and biexciton generation as a function of $N_p$ and the corresponding number of control parameters, $N_\theta$. Increasing the sequence from two to three pulses yields a noticeable improvement in fidelity. This initial gain is more pronounced for the bi-exciton case, as traversing the three-level ladder system benefits significantly more from the extra control degrees of freedom compared to the simpler two-level single-exciton transitions. However, for both systems, increasing the sequence beyond three pulses results in strictly diminishing returns. 

The underlying reason for this saturation becomes clear upon inspecting the optimized driving fields (detailed in Appendix~\ref{appendix:Parameter_Data.}): once an optimal low-pulse protocol is established, the optimizer tends to suppress or merge the additional pulses, effectively recovering the dynamics of the more fundamental two- or three-pulse sequences. Consequently, as indicated by the plateau in Fig.~\ref{fig:Multi-pulse-plot} for large pulse numbers, we conclude that the optimization reaches an effective operational limit dictated by the interplay of the phonon bath and the strict experimental constraints we have imposed, rather than being limited by the available control parameters. Given the significant experimental overhead required to synthesize and perfectly align extended pulse sequences, these marginal theoretical gains do not justify implementing schemes beyond the highly efficient two-pulse approach.

\subsection{Robustness to temperature sensitivity}
\label{sec:temp_dep}

Increasing the initial bath temperature enhances the thermal phonon population, which naturally strengthens the system-environment interaction 
due to electron-phonon scattering processes occuring during the excitation process. 
This increases phonon induced dephasing mechanisms \cite{Nazir_2016}, and consequently reduces the fidelity of population inversion. 
Here, we evaluate the temperature resilience of our optimized SUPER and FTPE protocols against standard $\pi$-pulse and two-photon excitation (TPE) schemes, and assess whether incorporating chirped pulses can further mitigate excitation induced dephasing.

\begin{figure}[t!]
\centering
\hspace{-1cm}
    \includegraphics[width=0.45\textwidth]{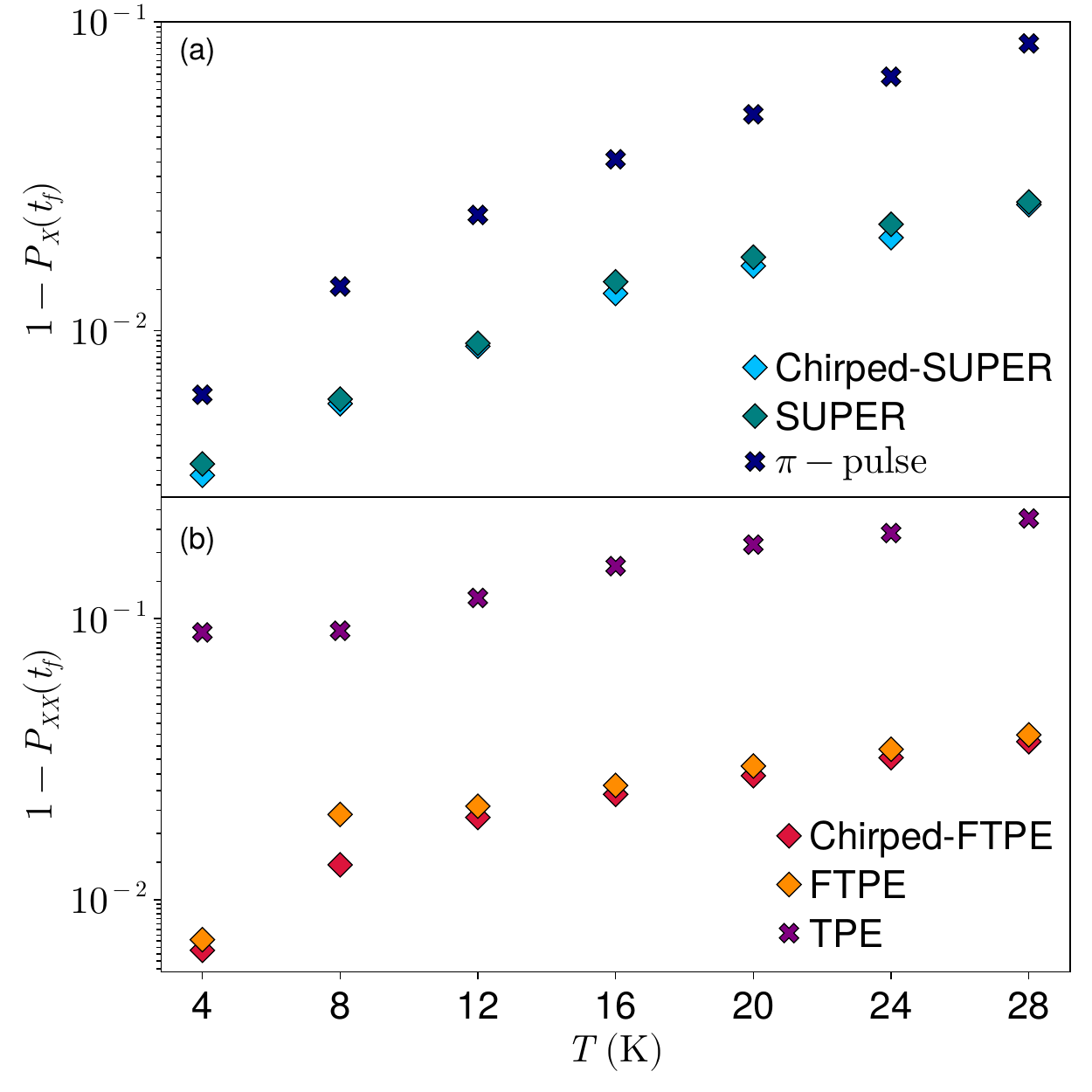}
    \caption{Temperature dependence of the exciton (a) and biexciton (b) infidelities, $1-P_{\Psi}(t_f)$, on a logarithmic scale for the indicated pulse protocols, with individually optimised parameters.} 
    \label{fig:Temp_plot}
\end{figure}

As illustrated in Fig.~\ref{fig:Temp_plot}, the optimized SUPER and FTPE protocols demonstrate significantly greater robustness to thermal fluctuations than their conventional resonant counterparts. This performance advantage becomes particularly striking at elevated temperatures. At 28 K, we observe a population inversion of approximately 97\% using the SUPER scheme, compared to just 91\% achieved under conventional $\pi$-pulse excitation. The improvement is even more pronounced for biexciton generation: while the standard TPE fidelity collapses to 78\% at 28 K, the optimized FTPE scheme preserves a remarkable 96\% fidelity.
Furthermore, the incorporation of linear frequency chirp produces a modest, albeit not entirely consistent, enhancement in preparation fidelity for both target states across the evaluated temperature range. By instantaneously sweeping the driving frequency, chirped pulses reduce the system's susceptibility to any single resonant phonon mode, effectively averaging out phonon-induced dephasing. While frequency-chirped pulses are well established for achieving robust population inversion and reducing phonon sensitivity in standard single-pulse setups \cite{PhysRevLett.68.2000, Malinovsky2001, PhysRevB.95.241306, PhysRevLett.106.166801, PhysRevLett.106.067401}, our results confirm that integrating chirp into multi-pulse SUPER and FTPE schemes provides an consistent improvement, though with diminishing returns for more complex pulse sequences.
\section{Conclusion}
We have introduced a highly efficient numerical optimization framework for strongly coupled, non-Markovian open quantum systems by integrating the uniTEMPO process tensor approach~\cite{link2024open} with automatic differentiation~\cite{baydin2018automaticdifferentiationmachinelearning, 1995SJSC...16.1190B}. Building on the efficient backpropagation frameworks established in~\cite{Butler2023, OrtegaTaberner2024ProcessTensor}, our approach implements gradients at a lower level to avoid the manual finite-difference calculation of propagator derivatives. This refinement eliminates a key computational bottleneck, likely accounting for the significant speed-up observed in our results; a detailed comparison of these methodologies is provided in Appendix~\ref{appendix:Comparison}.

To demonstrate the utility of this framework, we optimized multi-pulse SUPER~\cite{PRXQuantum.2.040354} and FTPE~\cite{yan2025robustentangledphotongeneration} excitation schemes for single- and biexciton preparation in semiconductor quantum dots. Constraining the control fields to strictly experimentally feasible regimes at an initial bath temperature of $4\text{K}$, our solver identified protocols yielding preparation fidelities of $99.63\%$ and $99.29\%$, respectively. By expanding the optimization space, we further demonstrated that while increasing the sequence to three pulses offers marginal fidelity gains, the system dynamics rapidly saturate. Beyond three pulses, the protocols encounter diminishing returns bounded by the fundamental limits of the phonon bath, confirming the practical optimality of low-pulse sequences.

Crucially, our framework addresses a major bottleneck in the experimental implementation of SUPER and FTPE protocols. While these schemes offer promising alternatives to resonant excitation~\cite{Stievater2001, PhysRevLett.87.246401}, the high-dimensional parameter space—even for a simple two-pulse sequence—is prohibitively large to explore either experimentally or via traditional master-equation simulations. Consequently, previous studies have often fixed specific pulse parameters to retain computational and experimental feasibility, potentially restricting the results to local maxima determined by these initial assumptions~\cite{PhysRevB.107.195306, Karli2022SUPER, yan2025robustentangledphotongeneration, PhysRevB.109.245304, Torun2026}. Using our optimisation scheme, we efficiently probe the full parameter landscape within practical experimental bounds. This approach identifies globally optimal solutions rather than relying on restricted parameter scans. We demonstrate that near-unity fidelity does not require a blind search over pulse energies; instead, our framework systematically determines precise, globally optimised configurations that reconcile high-fidelity operation with experimentally realistic constraints. Furthermore, these optimized multi-pulse sequences exhibit significantly greater robustness to elevated bath temperatures compared to standard $\pi$-pulse and two-photon excitation schemes. We also demonstrated that incorporating adiabatic linear chirp further insulates the system against phonon-induced dephasing. 

While this work focused on maximizing a single-objective, final-time state fidelity, the utility of the AD-uniTEMPO framework extends far beyond this specific application. Because the automated exact gradients naturally accommodate time-dependent and multi-objective cost functions, future work will leverage this capability to explore complex thermodynamic phenomena. While uniTEMPO is used here as the most efficient algorithm for simulating stationary bath dynamics, its selection is not a strict requirement for this numerical optimization framework. Crucially, AD can be applied to any algorithm utilizing the process tensor for open quantum evolution, such as PT-TEMPO~\cite{strathearn2018efficient, Jorgenson2019} for non-stationary baths or ACE~\cite{CygorekACE2024} for composite environments.
Ultimately, this work provides a versatile, computationally tractable toolset for the design of realistic quantum control protocols, paving the way for more robust solid-state quantum technologies.

\vspace{1em}
\section*{Acknowledgements}

SM acknowledges support from the EPSRC (EP/W524347/1). WB acknowledges support from the EPSRC (EP/T517793/1 and EP/W524335/1). HJDM acknowledges funding from a Royal Society Research Fellowship (URF/R1/231394). TJE is supported by the University of Manchester Dame Kathleen Ollerenshaw Fellowship. A.J.B. acknowledges additional support from the EPSRC Quantum Technologies Fellowship EP/W027909/1. JIS acknowledges support from the EPSRC (UKRI2863). We thank Ahsan Nazir for valuable discussions. 

\appendix

\section{Process Tensor Calculations}
\label{appendix:uniTEMPO_algorithm}

At the core of our approach is the uniTEMPO algorithm~\cite{link2024open}. In this Appendix we provide an overview of how process tensors are computed using uniTEMPO and explain why it is well suited to the task of optimizing control protocols that govern the system's dynamics. The time-evolving matrix product operator (TEMPO) method is a tensor network formulation of the discrete-time path integral approach—specifically, the quasi-adiabatic path integral (QuAPI)~\cite{Makri1995a, Makri1995b}. The tensor network techniques use a controlled truncation of information via singular value decomposition (SVD), permitting a low-rank representation of the discretised environment influence. Successive improvements on the TEMPO method have greatly reduced the computational complexity of such calculations~\cite{Cygorek2024,link2024open}.

The Trotter splitting of the total Hamiltonian $H(t)$ separates the components of the Hamiltonian and the influence term needs only be computed for each coupling and temperature once~\cite{jorgensenExploitingCausalTensor2019}. With respect to a time-dependent Hamiltonian, the evolution of a closed quantum system from time $t_0$ to time $t$ is given by $\rho(t) = U(t,t_0)\rho(t_0)U^\dagger(t,t_0)$ where the unitary operator takes the form
\begin{equation}
    U(t,t_{0}) = \mathcal{T}\exp\left(-i\int_{t_{0}}^{t}ds\,H(s)\right)
    \;,
\end{equation}
where $\mathcal{T}$ indicates chronological time-ordering. Rewriting this in terms of the superoperator $\mathcal{U}(t,t_0)$, the density operator is then $\rho(t)=\mathcal{U}(t,t_0)[\rho(t_0)]$. In Liouville space~\cite{Gyamfi_2020}, density matrices are represented as $d^2$-dimensional vectors and superoperators are represented by $d^2\times d^2$ matrices.  

In terms of the small homogenous time-step, $\delta t$, we find that $\rho(t_0+\delta t)\approx \exp(-iH(t_0)\delta t)\rho(t_0)\exp(iH(t_0)\delta t)$. Therefore, the total evolution from $t_0=0$ to $t$ is approximated by $U(t,0) \approx \prod_{i=0}^{N-1}\exp(-iH(t_i)\delta t)$ where $t=N\delta t$.

For each exponential in the dynamics, we apply a symmetric Trotter splitting to obtain,
\begin{equation}
    \begin{split}
        e^{-iH(t)\delta t} = e^{-iH_S(t)\delta t/2}e^{-i(H_E + H_I)\delta t}e^{-iH_S(t)\delta t/2} \\ 
        + \mathcal{O}(\delta t^3)
    \end{split}
\end{equation}
for which we have now separated the system Hamiltonian from that of the environment and its influence.

\begin{figure}[h]
    \centering
    \includegraphics{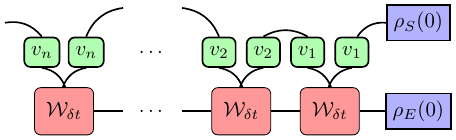}
    \caption{Discretisation of the open quantum evolution in terms of the superoperators $\mathcal{W}_{\delta t}$, implementing the time evolution generated by $H_I + H_E$, and $v_i=\mathcal{V}_{\delta t}^{1/2}(t_i)$ generated by $H_S(t_i)$ acting on an initial product state $\rho_S(0)\otimes\rho_E(0)$.}
    \label{fig:process-tensor}
\end{figure}

In terms of superoperators, we can write the time evolution as $\mathcal{U}(t+\delta t,t) \approx \mathcal{V}_{\delta t}^{1/2}(t)\mathcal{W}_{\delta t}\mathcal{V}_{\delta t}^{1/2}(t)$ where $\mathcal{V}_{\delta t}(t)[\rho]=e^{-iH_S(t)\delta t}\rho e^{iH_S(t)\delta t}$ is the system propagator, whereas $\mathcal{W}_{\delta t}[\rho] = e^{-i(H_E + H_I)\delta t}\rho e^{i(H_E + H_I)\delta t}$ which acts on both the system and environment. The product $\prod_{i=0}^{N-1}\mathcal{V}_{\delta t}^{1/2}(t_i)\mathcal{W}_{\delta t}\mathcal{V}_{\delta t}^{1/2}(t_i)$ can then be viewed in tensor network notation as a process tensor according to Fig.~\ref{fig:process-tensor}.

The infinite-dimensional Hilbert space of the environment makes direct calculation of $\mathcal{W}_{\delta t}$ impossible. However, the discretised Feynman-Vernon path integral can be used in conjunction with compression via singular-value decomposition to obtain a low-rank representation with numerically exact (convergent) results. The uniTEMPO~\cite{link2024open,kahlertSimulatingLandauZener2024,sonnerSemigroupInfluenceMatrices2025} method is a state-of-the-art matrix product state (MPS) technique for the computation of the influence functional using a novel diagonal contraction scheme which recasts the TEMPO contraction scheme as time-evolving block decimation (TEBD)~\cite{vidalEfficientSimulationOneDimensional2004}. A further improvement was identified using uniform matrix product states (uMPS) which emerge as a result of the stationarity of the initial state of the bath. The algorithm is then based on infinite time-evolving block decimation (iTEBD)~\cite{vidalClassicalSimulationInfiniteSize2007}. This also removes the simulation time as an input parameter. When using finite MPS methods, the size of the MPS limits the maximum duration of a simulation that can be carried out. With uniTEMPO, arbitrary time calculations can be carried out. The computational complexity of calculating the environment influence, in terms of the number of SVD operations, drastically reduces to $\mathcal{O}(N_C)$ from $\mathcal{O}(NN_C)$, where $N$ is the number of simulation time steps and $N_C$ is the number of time steps of bath memory resulting from the finite memory approximation~\cite{strathearnEfficientRealtimePath2017}.

In the present work, this largely simplifies the calculations where temperature is varied. We do not consider temperature as an optimisation parameter as this would bring in the computation of the influence into the automatic differentiation system, which would add significiant complexity.

\begin{figure}[h]
    \centering
    \includegraphics{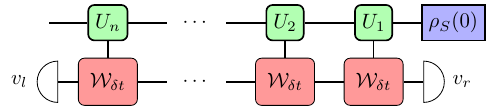}
    \caption{Calculation of the final state $\rho_S(t_n)$ after $n$ time steps. The initial state of the environment is incorporated into the uniTEMPO algorithm resulting in the fixed points $v_l$ and $v_r$ which are created to truncate the evolution after chosen simulation time. The pairs of unitaries $v_i$ are combined to form the rank-3 tensors $U_i$.} 
    \label{fig:final-state-tn}
\end{figure}

The process tensor can then be obtained for any protocol realisation. We keep this in a tensor network form to ease the automatic differentiation system. From here, calculations of observables, such as our objective function, are carried out by tensor contraction [Fig.~\ref{fig:final-state-tn}]. 

\section{Automatic Differentiation and Optimisation}
\label{appendix:Auto_diff}

Automatic differentiation~\cite{baydin2018automaticdifferentiationmachinelearning, 1995SJSC...16.1190B} is a widely used technique in nonlinear optimization, machine learning, and sensitivity analysis, providing a highly efficient and accurate method for computing the partial derivatives of a function specified by a computer program. Automatic differentiation exploits the fact that every computer calculation, executes a sequence of 
elementary arithmetic operations and elementary functions. By applying the chain rule repeatedly to these operations, partial derivatives of arbitrary order can be computed automatically, accurately to working precision, and using at most a small constant factor of additional arithmetic operations than the original program. Fundamental to automatic differentiation is the chain rule, which decomposes the derivatives. There exists two distinct types of automatic differentiation these are namely forward mode \cite{10.1145/3716309} and reverse mode \cite{aehle2022reversemodeautomaticdifferentiationcompiled}. Forward mode propagates derivatives alongside the primal evaluation, moving from inputs to outputs. For a computational graph with intermediate variables $w_1, w_2, \dots, w_n$ and an output $y = w_n$, the forward mode tracks the ``tangent'' $\dot{w}_i$ with respect to an input $x$:
\begin{equation}
\dot{w}_i = \frac{\partial w_i}{\partial x} = \frac{\partial w_i}{\partial w_{i-1}} \dot{w}_{i-1},
\end{equation}
where $\dot{w}_0 = 1$. In contrast, reverse mode (often referred to as backpropagation) propagates derivatives from the output back to the inputs. This mode computes the so-called adjoint $\bar{w}_i$, which represents the sensitivity of the final output $y$ to changes in the intermediate variable $w_i$:
\begin{equation}
\bar{w}_i = \frac{\partial y}{\partial w_i} = \bar{w}_{i+1} \frac{\partial w_{i+1}}{\partial w_i},
\end{equation}
where the process is initialized with the seed $\bar{y} = \bar{w}_n = 1$. The relative efficiency of these modes depends on the dimensions of the function's input and output spaces. Forward mode is more computationally efficient for functions mapping a small number of inputs to a large number of outputs. Conversely, reverse mode is significantly more performant for scalar-valued objective functions where the number of input control parameters is large. Given that our optimization involves up to 23 independent control parameters but only a single output value (the target state fidelity), reverse mode is the more suitable choice. 

In reverse mode, the output variable is held fixed while derivatives with respect to intermediate subexpressions are computed recursively. The chain rule is applied to propagate sensitivities backward through the computational graph, efficiently accumulating the gradient with respect to the input variables.
\begin{equation}
\frac{\partial y}{\partial x}
= \frac{\partial y}{\partial w_1}\frac{\partial w_1}{\partial x}
= \left(\frac{\partial y}{\partial w_2}\frac{\partial w_2}{\partial w_1}\right)\frac{\partial w_1}{\partial x}
= \dots
\end{equation}
Here $w_1$ and $w_2$ represent intermediate variables of the overall computational graph. This is in contrast to forward mode, where one first fixes the independent variable with respect to which differentiation is performed and computes the derivative of each sub-expression recursively. The adjoint is a derivative of a chosen dependent variable with respect to a subexpression $w_i$ and can be written as
\begin{equation}
\bar{w_i} = \frac{\partial{y}}{\partial{w_i}}.
\end{equation}
Using the chain rule if $w_i$ has successors in the computational graph, we have
\begin{equation}
\bar{w}_i = \sum_{j \in \text{successors of i}} \bar{w}_j \frac{\partial w_j}{\partial w_i}
\end{equation}

\section{Computational Details}
\label{appendix:Computaional_details}

The uniTEMPO method and the tensor contractions to calculate observables are implemented in the Julia programming~\cite{bezansonJuliaFastDynamic2012, doi:10.1137/141000671}. Unlike in Python, AD implementations are generally native in Julia. The multiple dispatch paradigm used Julia exploits is exploited by AD. For example, for forward-mode, the same source code can be used to define the objective function and gradient. Julia just-in-time compiles a method using dual-numbers based on your original function definition. This has the benefit of removing the need to rewrite your functions to be AD compatible. For reverse-mode AD, Julia's support for meta-programming allows the AD machinery to examine the abstract syntax tree of your source code to determine the gradient. Finally, Julia has a strong ecosystem in tensor networks and scientific machine learning, making it the ideal choice for combining the two. In our situation, we used the Zygote.jl~\cite{innes2019dontunrolladjointdifferentiating} package (a source-to-source reverse-mode implementation). However, it is trivial to switch between forward-mode, reverse-mode and finite difference techniques~\cite{dalle2025commoninterfaceautomaticdifferentiation}. Furthermore, it is possible to compute the Hessian for optimisers that require higher-order derivatives. Lastly, we note that acceleration via graphical processing units is a possibility. We had no need to exploit it in the current work, however, we are considering it for more complex optimisation problems for future works. 

To initialize the optimization scheme and avoid entrapment in local optima we implemented a pre-optimization phase. Specifically, we employed the random-search algorithm from the BlackBoxOptim.jl library~\cite{vaibhav_kumar_dixit_2023_7738525} to broadly sample the parameter space. The implementation of the gradient-based optimisation uses the L-BFGS algorithm from the Optim.jl, library~\cite{Mogensen2018}. While various optimization algorithms are available, we employ L-BFGS in favour of first-order methods such as Adam or standard gradient descent, as it demonstrated superior convergence properties for this specific control problem. To ensure our control pulses remain within physically realistic limits, we utilize a box-constrained minimization. This approach enforces upper and lower bounds on the pulse parameters while navigating the non-linear landscape of the fidelity objective function. The solver is configured with two outer iterations, each containing 100 inner iterations. The L-BFGS algorithm used a memory limit of $m = 10$, an initial step size of $\alpha_0 = 1.0$ and the line search was conducted using the Moré-Thuente method. In general, when using uniTEMPO, we employed a time-step of $\delta t = 0.01$ ps and a truncation tolerance of $\Delta = 1 \times 10^{-10}$ for the singular value decomposition.

Numerical simulations for the two-pulse sequences—including both standard and chirped configurations across the investigated temperature range—were performed locally on an Intel Core i7-12700 processor (12 cores, up to 4.90 GHz). For more computationally intensive tasks, specifically the multi-pulse sequences and the statistical averaging of fidelity over multiple runs, we utilized the Computational Shared Facility (CSF3) at the University of Manchester. These simulations were executed on standard compute nodes equipped with AMD EPYC Genoa 9634 CPUs (2.25 GHz), typically utilising 4 cores per job. Computational runtime for the standard two-pulse optimization at $T=4$~K is approximately 30 minutes. While the complexity increases with the addition of more pulses or the introduction of a frequency chirp, the total runtime does not scale significantly, remaining within a manageable range for all investigated protocols.

\section{The Cost Function Penalty.}
\label{appendix:The Cost Function}
Here we introduce the penalty term in the cost function to restrict spectral overlap with the zero-phonon line (ZPL), thereby ensuring that the pulses considered remain genuinely off-resonant. The penalty is defined as
\begin{equation}
\mathcal{P} = \lambda\,\Theta\!\left(R - R_c\right),
\label{eq:Penalty_func}
\end{equation}
where $\lambda$ is the relative weighting factor, which we choose to be $\lambda = 100$, and $\Theta$ denotes the Heaviside step function. The quantity $R$ represents the relative spectral overlap with the ZPL, and $R_c$ is the threshold value, chosen here to be $10\%$. This penalty term is subtracted from the original cost function \eqref{eq:Cost_func_general}, ensuring that regions of parameter space corresponding to more than $10\%$ ZPL overlap are effectively ignored during the optimisation process.

To calculate the relative spectral overlap with the ZPL, we take the Fourier transform of the pulse envelope being used. For the standard fixed-detuning case, this corresponds to the Fourier transform of Eq.~\eqref{eq:Pulse_envelope}, and for chirped pulses, it corresponds to the Fourier transform of Eq.~\eqref{eq:Chirp_envelope}. The spectral power is then obtained as the modulus squared of the Fourier transform. The fraction of the pulse that overlaps with the ZPL is estimated by considering the distribution of pulse energy across frequencies, the ratio of energy in the undesired spectral region to the total pulse energy gives a direct measure of overlap. Specifically, for SUPER-like protocols, only negative-frequency components are allowed, since we typically employ red-detuned lasers. Any pulse energy in the positive-frequency region is therefore treated as overlapping with the ZPL and contributes to spectral leakage. For FTPE-like protocols, the situation is slightly more complex, as detuning can occur on both sides of the spectrum. Nevertheless, the procedure is similar: we integrate the pulse energy in the undesired frequency region (e.g., negative frequencies if the pulse is positively detuned) and normalize by the total pulse energy. This gives a reasonably good estimate of $R$ that can be used in Eq.~\eqref{eq:Penalty_func}.
\section{Parameter Data}
\label{appendix:Parameter_Data.}
Here we provide all parameter values obtained through our automatic differentiation optimisation routine as presented in the main text. Table~\ref{tab:combined_parameters} lists the optimal parameters identified through our search, representing the values that yielded the maximum state fidelity for an increasing number of pulses. These results are used to obtain the average state fidelity as is presented in Figure~\ref{fig:Multi-pulse-plot}. 
\begin{table*}[t!]
\centering
\small
\renewcommand{\arraystretch}{1.2} 
\setlength{\tabcolsep}{3pt} 
\caption{Optimized parameters for exciton preparation via the SUPER scheme and biexciton preparation via the FTPE scheme. Values are obtained via automatic differentiation for a number of pulses varying from $n=2$ to $n=6$.}
\label{tab:combined_parameters}
\begin{tabular*}{\textwidth}{@{\extracolsep{\fill}}l cccc cccc}
\toprule
& \multicolumn{4}{c}{\textbf{Exciton Parameters (SUPER)}} & \multicolumn{4}{c}{\textbf{Biexciton Parameters (FTPE)}} \\
\cmidrule(lr){2-5} \cmidrule(lr){6-9}
Pulse & $\delta$ (ps$^{-1}$) & $\sigma$ (ps) & $\Theta (\pi)$ & $\tau$ (ps) & $\delta$ (ps$^{-1}$) & $\sigma$ (ps) & $\Theta(\pi)$ & $\tau$ (ps) \\
\midrule
\multicolumn{9}{c}{\textbf{Number of pulses = 2}} \\
\midrule
1 & $-5.36$ & $1.00$ & $11.84$ & --     & $-13.77$ & $1.00$ & $14.00$ & -- \\
2 & $-20.00$ & $1.00$ & $8.44$  & $1.00$   & $13.93$ & $1.00$ & $14.04$  & $-0.15$ \\
\midrule
\multicolumn{9}{c}{\textbf{Number of pulses = 3}} \\
\midrule
1 & $-11.28$ & $15.25$ & $6.48$ & --     & $-14.90$ & $2.91$ & $6.97$ & -- \\
2 & $-15.17$  & $1.00$ & $12.00$ & $0.89$ & $12.28$  & $1.00$ & $14.80$ & $-1.55$ \\
3 & $-2.71$ & $1.00$ & $11.30$  & $-0.90$ & $-11.72$ & $1.00$ & $15.00$  & $-1.29$ \\
\midrule
\multicolumn{9}{c}{\textbf{Number of pulses = 4}} \\
\midrule
1 & $-16.15$  & $7.54$ & $8.92$  & --     & $-14.30$  & $4.94$ & $5.37$  & -- \\
2 & $-6.09$ & $16.64$  & $1.53$ & $-0.55$   & $14.73$ & $4.62$  & $9.97$ & $-1.93$ \\
3 & $-17.56$  & $1.00$  & $11.65$  & $1.00$ & $-12.54$  & $1.00$  & $13.43$  & $-0.45$ \\
4 & $-3.21$ & $1.00$  & $12.00$  & $-0.73$   & $11.64$ & $1.00$  & $14.40$  & $-0.68$ \\
\midrule
\multicolumn{9}{c}{\textbf{Number of pulses = 5}} \\
\midrule
1 & $-9.43$ & $18.24$  & $8.93$  & --     & $-14.54$ & $9.99$  & $9.81$  & -- \\
2 & $-2.04$ & $1.00$ & $5.60$  & $-0.71$ & $12.47$ & $2.09$ & $12.39$  & $0.89$ \\
3 & $-4.66$ & $1.32$  & $10.52$  & $-0.90$   & $-3.00$ & $1.00$  & $5.60$  & $1.47$ \\
4 & $-16.78$ & $2.28$  & $9.70$ & $1.00$ & $11.86$  & $1.58$  & $13.53$ & $0.15$ \\
5 & $-19.69$ & $1.00$  & $10.10$  & $0.96$   & $-13.14$ & $5.11$  & $3.63$  & $0.59$ \\
\midrule
\multicolumn{9}{c}{\textbf{Number of pulses = 6}} \\
\midrule
1 & $-15.67$ & $13.54$ & $6.21$ & --      & $-10.13$ & $1.23$ & $4.01$ & -- \\
2 & $-11.00$ & $18.75$ & $7.14$ & $-0.15$  & $12.31$ & $1.00$ & $12.79$ & $-0.09$ \\
3 & $-13.32$  & $8.24$ & $6.83$ & $0.26$  & $-13.03$  & $1.28$ & $7.09$ & $0.69$ \\
4 & $-3.84$  & $1.26$ & $10.02$  & $-0.64$  & $9.58$  & $1.77$ & $3.02$  & $-1.99$ \\
5 & $-20.00$ & $1.00$ & $10.12$  & $1.00$   & $-11.44$ & $1.00$ & $8.14$  & $-0.58$ \\
6 & $-4.97$  & $1.29$ & $7.57$ & $-0.12$   & $10.21$  & $6.00$ & $7.19$ & $-1.55$ \\
\bottomrule
\end{tabular*}
\end{table*}
The table includes the pulse detuning $\delta$ (ps$^{-1}$), pulse width $\sigma$ (ps), pulse area $\Theta$ (in units of $\pi$), and the time delay relative to the first pulse, $\tau$ (ps). 
For single-exciton preparation, the optimization constraints were defined as $\delta \in [-20, -1]~\mathrm{ps}^{-1}$,
$\sigma \in [1, 20]~\mathrm{ps}$, $\tau \in [-1, 1]~\mathrm{ps}$, and $\Theta \in [0.5\pi, 12\pi]$.
For biexciton generation, the ranges for the pulse area and delay were defined as $\Theta \in [2.5\pi, 15\pi]$
$\tau \in [-2, 2]~\mathrm{ps}$ while maintaining
$\sigma \in [1, 20]~\mathrm{ps}$.
The detuning for the biexciton case ($\delta$) was implemented using a two-color scheme; specifically, positive detuning was restricted to the interval $[3, 15]~\text{ps}^{-1}$, while negative detuning was restricted to $[-15, -3]~\text{ps}^{-1}$. Our results indicate that for pulse sequences with an even number of pulses, a symmetric distribution of positive and negative detunings is optimal. For sequences with an odd number of pulses, however, the sign of the detuning for the additional pulse had no significant impact on the final fidelity. 
The choice to employ only negative detuning for single-exciton preparation is motivated by the protocol used in the SUPER scheme, which we adopt as a benchmark. In our investigations, this approach was also found to perform more reliably than a two-colour scheme for exciton preparation. A similar consideration applies to biexciton preparation. FTPE employs a two-colour scheme, and in our simulations this approach consistently outperformed schemes restricted to either exclusively negative or exclusively positive detuning. Accordingly, our choices of detuning configurations for both single- and bi-exciton preparation are guided by the performance of these pre-existing schemes. The ranges for detuning and pulse width were chosen to minimize overlap with the zero-phonon line (ZPL), while the lower bound on pulse width ensures that the required optical setup remains feasible. The upper bound of the pulse area ensures that degrees of freedom in the bulk semiconductor are not inadvertently excited, which would lead to increased dephasing.
The optimized pulse widths $\sigma$ are generally small (\mbox{$\sim$1--3~ps}), while the pulse areas $\Theta$ are relatively large, typically $5$--$12\pi$ for exciton and $5$--$15\pi$ for biexciton generation. Smaller $\sigma$ provides precise excitation with minimal unwanted transitions, and larger $\Theta$ ensures sufficient Rabi rotation to maximize population inversion.
For single-exciton generation, our control strategy is motivated by the SUPER scheme~\cite{PRXQuantum.2.040354}. We find that the optimized detunings $\delta$ typically converge toward two distinct regimes, approximately $-20$~ps$^{-1}$ and $-5$~ps$^{-1}$. Selecting these specific values for each pulse in the sequence is essential to establish the constructive interference required for high-fidelity population inversion.
\begin{table*}[t!]
\centering
\small
\renewcommand{\arraystretch}{1.2} 
\setlength{\tabcolsep}{2pt} 
\caption{Optimized parameters for exciton preparation via SUPER and biexciton preparation via FTPE versus initial bath temperature $T$. Each sequence comprises two pulses separated by delay $\tau$. Units: $\tau, \sigma$ (ps), $\delta$ (ps$^{-1}$), and $\Theta$ ($\pi$).}
\label{tab:temp_biexciton_combined}
\begin{tabular*}{\textwidth}{@{\extracolsep{\fill}}l ccccccc ccccccc}
\toprule
& \multicolumn{7}{c}{\textbf{Exciton Parameters}} & \multicolumn{7}{c}{\textbf{Biexciton Parameters}} \\
\cmidrule(lr){2-8} \cmidrule(lr){9-15}
$T$ (K) & $\tau$ & $\Theta_1$ & $\Theta_2$ & $\delta_1$ & $\delta_2$ & $\sigma_1$ & $\sigma_2$ & $\tau$ & $\Theta_1$ & $\Theta_2$ & $\delta_1$ & $\delta_2$ & $\sigma_1$ & $\sigma_2$ \\
\midrule
4  &  1.00 & 11.84 & 8.44 & -5.36 & -20.00 & 1.00 & 1.00 & -0.15 & 14.00 & 14.03 & -13.77 & 13.93 & 1.00 & 1.00 \\
8  & -1.00 & 8.24 & 11.96 & -20.00 & -5.30 & 1.00 & 1.00 & -0.17 & 13.33 & 13.47 & -12.53 & 12.44 & 1.00 & 1.00 \\
12 & -1.00 & 8.21 & 12.00 & -20.00 & -5.27 & 1.00 & 1.00 &  0.04 & 13.58 & 15.00 & -13.61 & 8.30 & 1.00 & 1.00 \\
16 & -0.66 & 12.00 & 12.00 & -20.00 & -5.30 & 2.32 & 1.00 &  0.04 & 15.00 & 13.67 & -8.03 & 13.67 & 1.00 & 1.00 \\
20 &  1.00 & 12.00 & 8.64 & -5.26 & -20.00 & 1.00 & 1.19 &  0.03 & 13.63 & 15.00 & -13.58 & 8.23 & 1.00 & 1.00 \\
24 & -0.65 & 12.00 & 12.00 & -19.99 & -5.29 & 2.33 & 1.00 &  0.03 & 13.65 & 15.00 & -13.58 & 8.22 & 1.00 & 1.00 \\
28 &  0.19 & 12.00 & 12.00 & -20.00 & -5.30 & 2.42 & 1.00 &  0.03 & 15.00 & 13.71 & -8.00 & 13.71 & 1.00 & 1.00 \\
\bottomrule
\end{tabular*}
\vspace{1ex}
\end{table*}
In contrast, for biexciton generation, we adapt the FTPE scheme~\cite{yan2025robustentangledphotongeneration}. Here, the optimization yields detunings of roughly equal magnitude but opposite sign. The optimized time delays $\tau$ show no apparent pattern and appear to serve primarily as fine-tuning parameters rather than variables of any significant impact on the overall dynamics.
\begin{table*}[t!]
\centering
\renewcommand{\arraystretch}{1.1} 
\setlength{\tabcolsep}{1.8pt}
\caption{Optimized parameters, including chirp rates ($\alpha$), for exciton preparation via SUPER and biexciton preparation via FTPE versus initial bath temperature $T$. Each sequence consists of two pulses with delay $\tau$, including a separate biexciton delay $\tau^b$. Units: $\tau, \sigma$ (ps), $\delta$ (ps$^{-1}$), $\Theta$ ($\pi$), and $\alpha$ (ps$^{-2}$).}
\label{tab:temp_chirp_biexciton}
\begin{tabular*}{\textwidth}{@{\extracolsep{\fill}} ll cccccccc c cccccccc}
\toprule
& & \multicolumn{8}{c}{\textbf{Exciton Parameters (SUPER)}} & & \multicolumn{8}{c}{\textbf{Biexciton Parameters (FTPE)}} \\
\cmidrule(lr){2-10} \cmidrule(lr){11-19}
$T$ & $\tau$ & $\Theta_1$ & $\Theta_2$ & $\delta_1$ & $\delta_2$ & $\sigma_1$ & $\sigma_2$ & $\alpha_1$ & $\alpha_2$ & $\tau$ & $\Theta_1$ & $\Theta_2$ & $\delta_1$ & $\delta_2$ & $\sigma_1$ & $\sigma_2$ & $\alpha_1$ & $\alpha_2$ \\
\midrule
4  & -1.00 & 7.86 & 12.00 & -20.00 & -5.66 & 1.00 & 1.00 & -0.12 & -0.33 & -0.14 & 14.87 & 14.03 & -13.55 & 14.58 & 1.00 & 1.00 & -0.44 & 0.00 \\
8  & -1.00 & 8.00 & 12.00 & -20.00 & -5.46 & 1.00 & 1.00 & -0.09 & -0.20 & 0.17  & 15.00 & 15.00 & -12.64 & 8.33  & 1.01 & 1.00 & 0.70  & 0.28 \\
12 & -0.42 & 10.36 & 9.82 & -7.09  & -2.53 & 1.00 & 1.00 & -0.05 & -0.05 & 0.13  & 15.00 & 15.00 & -8.56  & 12.63 & 1.00 & 1.00 & 0.00  & 0.69 \\
16 & 1.00  & 12.00 & 8.18 & -5.27  & -20.00 & 1.00 & 1.00 & -0.05 & -0.52 & 0.03  & 13.40 & 15.00 & -13.69 & 7.83  & 1.00 & 1.00 & 0.00  & 0.26 \\
20 & -1.00 & 8.59 & 12.00 & -20.00 & -5.31 & 1.23 & 1.00 & -0.05 & -0.05 & 0.10  & 15.00 & 14.80 & -7.96  & 12.79 & 1.00 & 1.00 & 0.26  & 0.70 \\
24 & -1.00 & 8.88 & 12.00 & -20.00 & -5.32 & 1.34 & 1.00 & -0.05 & -0.05 & 0.02  & 15.00 & 13.55 & -7.62  & 13.78 & 1.00 & 1.00 & 0.25 & 0.11 \\
28 & 1.00  & 12.00 & 12.00 & -5.31  & -20.00 & 1.00 & 2.14 & -0.05 & -0.70 & 0.02  & 13.53 & 15.00 & -13.64 & 7.87  & 1.00 & 1.00 & 0.09 & 0.23 \\
\bottomrule
\end{tabular*}
\end{table*}
An interesting feature of the optimization landscape is that as the number of pulses increases, the parameter values do not shift dramatically; instead, they largely recapitulate the behavior observed in two-pulse sequences. For example, in the three-pulse case for both exciton and biexciton generation, the pulse widths ($\sigma$) remain narrow, and the pulse areas ($\Theta$) remain high. Furthermore, the detunings ($\delta$) consistently adopt their characteristic values seen in the simpler sequences. This suggests that the fundamental control mechanism is established with two pulses, while additional pulses serve to refine the fidelity rather than qualitatively alter the underlying dynamics. This explains why upon increasing the number of pulses we observe diminishing returns.
Similarly, for Figure~\ref{fig:Temp_plot}, the optimized parameters from our optimization scheme are presented in Table~\ref{tab:temp_biexciton_combined}. As with the results shown in Figure~\ref{fig:Multi-pulse-plot}, we find that the pulse widths $\sigma$ tend to be small, while the pulse areas $\Theta$ are generally large. This again indicates that narrow, high-energy pulses are favourable. The observed parameter stability appears to have few exceptions; however, at elevated temperatures, the biexciton detunings occasionally deviate from the symmetric ($+\delta, -\delta$) pattern. Beyond this localized spectral shift, the optimized parameters at higher temperatures remain largely consistent with those found in the two-pulse case at $T=4\text{K}$. This indicates that the influence of phonon coupling does not significantly alter the dynamics of the optimised SUPER and FTPE schemes for single- and bi-exciton generation respectively. In other words, the optimal regime of the scheme is essentially preserved even under stronger phonon interactions, demonstrating the robustness of the control strategy against temperature-induced dephasing. Finally, we present the parameter data for the case of a chirped pulse, again for both single- and bi-exciton preparation at increased temperatures. The introduction of a linear frequency chirp modifies the pulse profile according to Eq.~\ref{eq:Chirp_envelope}, necessitating the inclusion of the chirp rate $\alpha$ as an additional degree of freedom in the optimization. For these simulations, we constrain the chirp rate to the range $\alpha \in [-0.70, 0.70]$~ps$^{-2}$. Since the instantaneous pulse width of a chirped pulse evolves temporally, we define the optimization constraint in terms of the transform-limited initial pulse width, $\sigma_0$, which we vary within the range $\sigma_0 \in [1, 10]$~ps. The range of initial pulse widths was decreased primarily because the optimal widths in the two-pulse protocol were consistently small. Furthermore, since pulse widths increase during the chirping protocol, maintaining the original constraints would include an unnecessarily large parameter space that is known to be sub-optimal.
 \section{Direct comparison to the PT-adjoint state method.}
\label{appendix:Comparison}
\begin{figure}[t!]
\centering
\vspace{1cm}
    \includegraphics[width=\columnwidth]{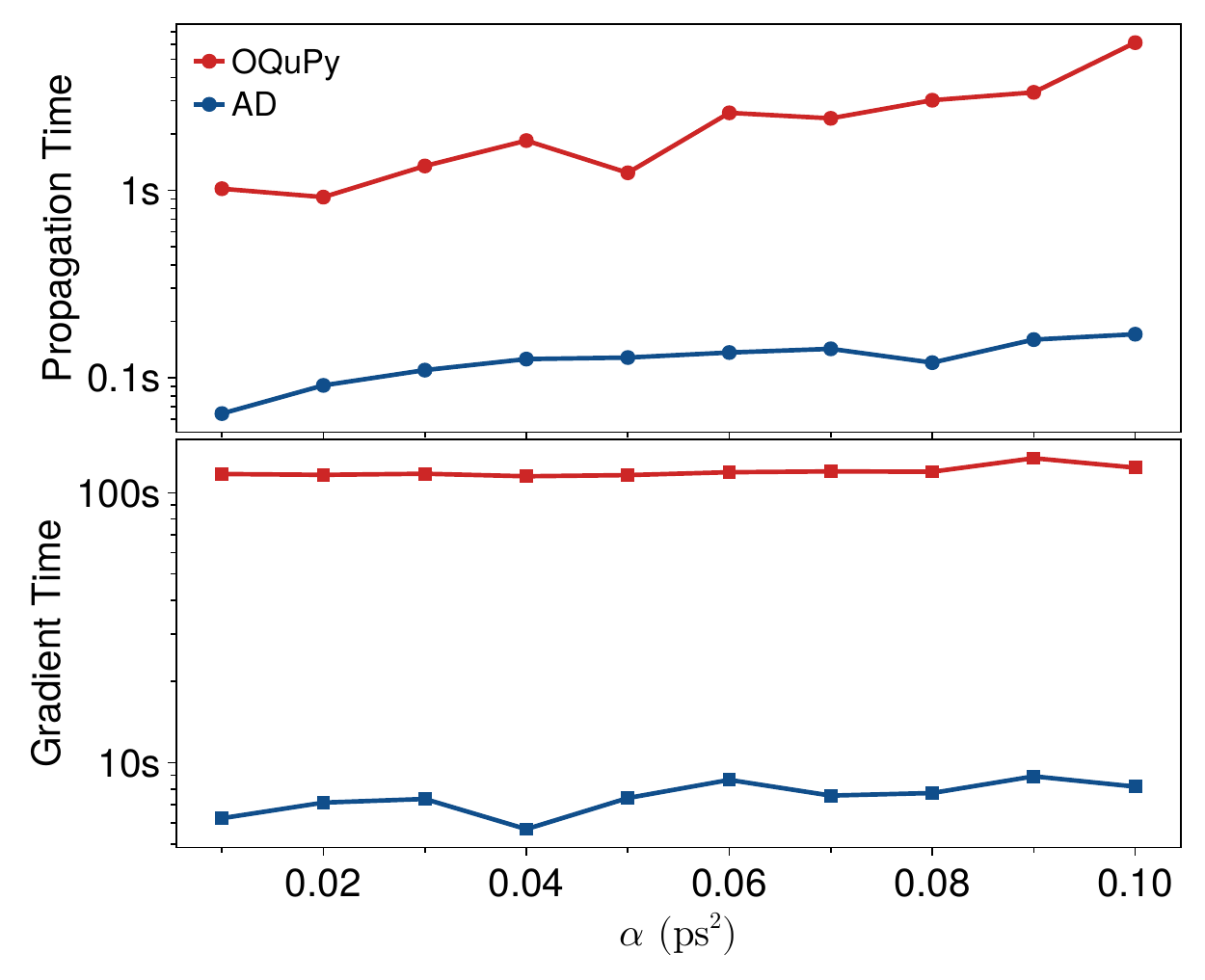}
    \caption{Benchmark comparing the AD method to the OQuPy adjoint state method. The top panel shows the time for a single forward propagation (cost-function evaluation), and the bottom panel shows the time to compute the gradient, both plotted as a function of the coupling strength $\alpha$.}
    \label{fig:benchmark}
\end{figure}

Now we detail a direct comparison between our method for optimisation and adjoint method of  \textcite{Butler2023,OrtegaTaberner2024ProcessTensor, ortegataberner2026qubitresetbornmarkovapproximation, ortegataberner2026quantumcontrolenvironmentopen}. To provide an accurate comparison, we focus on the optimization of the SUPER scheme in the standard two-pulse configuration. Detailed descriptions of the environment-system coupling and the underlying system dynamics are provided in Sections~\ref{sec:Phonon interactions} and~\ref{sec:Single-Excitons}, respectively. Figure \ref{fig:benchmark} illustrates the computation time for the gradient calculations and cost function evaluation for an increasing coupling strength. These simulations are performed with a time-step of $\delta t = 0.01$~ps. For each coupling strength, the SVD tolerance is individually tuned; we select the largest (coarsest) tolerance that maintains a relative error in the final exciton population of less than $0.01\%$, compared to a converged baseline calculation performed with a tolerance of $10^{-12}$.
Beyond the computational performance gains relative to previous works~\cite{Butler2023, OrtegaTaberner2024ProcessTensor}, our method offers significant practical and algorithmic advantages. By utilizing the uniTEMPO framework~\cite{link2024open}, the simulation duration does not need to be predefined; the temporal extent of the dynamics can be determined dynamically during the calculation. Furthermore, the integration of AD ensures that derivatives of the objective function are computed automatically. This removes the need for the manual derivation and implementation of adjoint equations required by the state-vector-based adjoint methods used in Refs.~\cite{Butler2023, OrtegaTaberner2024ProcessTensor}, making our approach adaptable to more complex observables and to other process tensor representations.
\section{Robustness to Pulse Phase Variations}
\label{appendix:phase_sensitivity}
\begin{figure}[h!]
\centering
\includegraphics[width=\columnwidth]{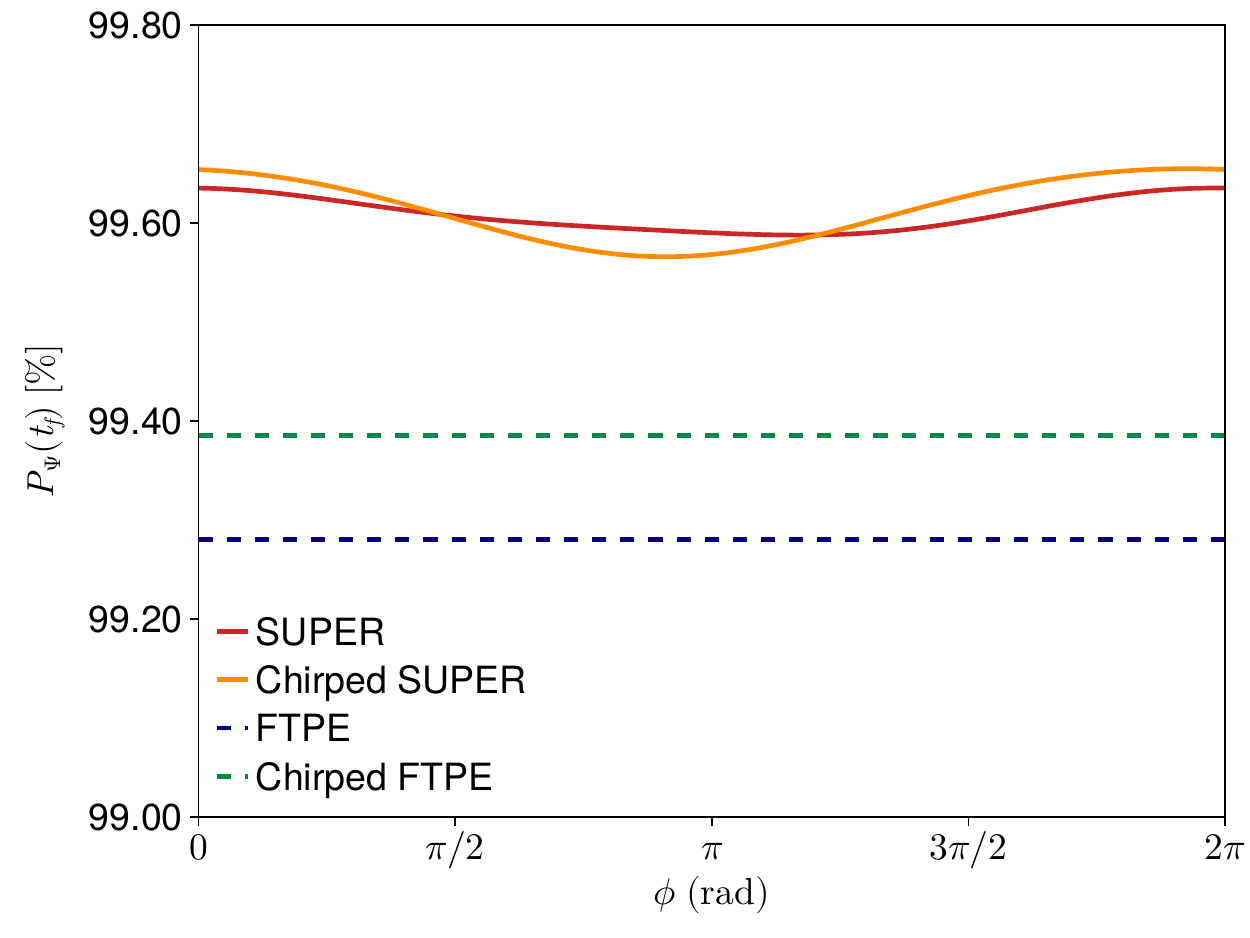}
\hspace{250em}
\vspace{-0.55cm}
\caption{
Dependence of the final excited state population on the relative phase between the two pulses for both single- and bi-exciton preparation in the chirped and unchirped cases. These have been plotted for the optimal parameter sets in all cases.}
\label{fig:Phase_dep}
\end{figure}
In the main text, we have assumed that the two Gaussian pulses are strictly phase-locked. While this simplifies the optimisation, achieving perfect phase coherence between pulses of different frequencies can be experimentally demanding. Although phase independence has been demonstrated in both the SUPER~\cite{PRXQuantum.2.040354} and FTPE~\cite{yan2025robustentangledphotongeneration} schemes, we explicitly verify this property for our optimized solutions. In particular, we investigate whether phase independence is maintained when chirping is included. To explore this, we introduce a relative phase $\phi$ to the second pulse
\begin{equation}
\Omega_2(t) \longrightarrow \Omega_2(t) e^{-i \phi}.
\label{eq: Phase dep}
\end{equation}
Our discussion of relative phase effects is restricted to two-pulse protocols, as these are the most experimentally viable. While three-pulse sequences offer improvements in the fidelity of both single- and bi-exciton state preparation, they present considerable experimental complexity. Consequently, analysing phase effects in the three-pulse case would offer little practical insight. Figure~\ref{fig:Phase_dep} presents the phase dependence of the preparation fidelity for the single-exciton and bi-exciton cases, respectively. 
The bi-exciton preparation fidelity is independent of the relative phase as previously reported~\cite{yan2025robustentangledphotongeneration}. In contrast, the single-exciton fidelity exhibits a slight dip, suggesting that the scheme is not entirely phase-independent, as had been previously suggested in Ref.~\cite{PRXQuantum.2.040354}. However, the variation is small ($< 0.1\%$), thus the scheme is still strongly robust to phase instability.

\bibliography{main}

\end{document}